\documentclass[nofootinbib,superscriptaddress]{revtex4-1}

\usepackage{amsmath}
\usepackage{amsfonts}
\usepackage{amssymb}
\usepackage[english]{babel}
\usepackage[latin1]{inputenc}
\usepackage{graphicx}
\usepackage{hyperref}
\usepackage{dsfont}
\usepackage[usenames,dvipsnames,svgnames]{xcolor}
\usepackage{float}

\begin{document}

\title{Investigation on Finsler geometry as a generalization to curved spacetime of Planck-scale-deformed relativity in the de Sitter case}

\newcommand{\addressZagreb}{Ru\dj er Bo\v{s}kovi\'{c} Institute, Division of Theoretical Physics, Bijeni\v{c}ka c.54, HR-10002 Zagreb, Croatia}
\newcommand{\addressRoma}{Dipartimento di Fisica, Universit\'{a} ``La Sapienza''
and Sez. Roma1 INFN, P.le A. Moro 2, 00185 Roma, Italy}
\newcommand{\addressICRA}{ICRANet, Piazza della Repubblica 10, I-65122 Pescara, Italy}
\newcommand{\addressCAPES}{CAPES Foundation, Ministry of Education of Brazil, Bras\'ilia, Brazil}

\author{Iarley P. Lobo}
\affiliation{\addressRoma}
\affiliation{\addressICRA}
\affiliation{\addressCAPES}
\author{Niccol\'{o} Loret}
\affiliation{\addressZagreb}
\author{Francisco Nettel}
\affiliation{\addressRoma}

\begin{abstract}
Over the last few years, Planck-scale modifications to particles' dispersion relation have been deeply studied for the possibility to formulate some phenomenology of Planckian effects in astrophysical and cosmological frameworks. There are some indications [arXiv:gr-qc/0611024] that Finsler geometry can provide some generalization of Riemannian geometry which may allow to account for non-trivial (Planckian) structure of relativistic particles' configuration space. We investigate the possibility to formalize Planck-scale deformations to relativistic models in curved spacetime, within the framework of Finsler geometry. We take into account the general strategy of analysis of dispersion relations modifications in curved spacetimes proposed in [arXiv:1507.02056], generalizing to the de Sitter case the results obtained in [arXiv:1407.8143], for deformed relativistic particle kinematics in flat spacetime using Finsler formalism.
\end{abstract}

\maketitle
\tableofcontents

\section{Introduction}
One of the most important definitions in Euclidean geometry concerns the norm of a vector and the distance between two points: they are defined using Pythagoras' theorem. In the XIX century, Riemann generalized  Gauss' ideas, introducing the concept of manifold and constructed the Riemannian geometry, in which Pythagoras theorem would be only valid at a point or along a line, but would no longer be the way of measuring distances in an open set in the manifold; instead, such measure would be performed using a general metric field. Euclidean geometry was the background of Newtonian mechanics and was invariant by Galilean boosts and rotations, but the advent of special relativity and a new kind of invariance group inspired the use of a geometry that unified space and time in a way that could keep invariant the interval between two events: such was Minkowski geometry, in which the interval represents a distance in such spacetime, that is measured by a formula similar to the Euclidean one, however allowing for positive, null or negative norms of vectors: this is a pseudo-Euclidean space. The next generalization, responsible for a relativistic description of gravity was performed by Einstein by writing General Relativity using a generalization of the Minkowski space by the same terms of the generalization from Euclidean to Riemannian geometry: spacetime was defined as a differential manifold endowed with a pseudo-Riemannian metric.
\par
In the geometries just described, one uses a metric to define the norm of vectors and co-vectors and, then defines the length of curves and the distance between points. An important feature of Special Relativity that gains a geometrical interpretation is the dispersion relation of particles, {\it id est} the equation that relates energy, momentum and mass, it is defined as the norm of the four-momentum in Minkowski spacetime. Therefore the pseudo-Euclidean nature of the Minkowski metric is what allows us to have a dispersion relation of the form $E^2-p^2=m^2$. Such relation can be generalized to a curved spacetime, where the dispersion relation is defined as the norm with the curved metric, and the Minkowskian form is achieved in normal coordinates.
\par
On the other hand, some approaches to quantum gravity phenomenology refers to the possibility of having a Modified Dispersion Relation (MDR) as a property of the semi-classical limit of quantum gravity theories, for example $m^2=E^2-p^2+\ell E^3$, where $\ell$ is a deformation parameter that depends on the model (which in some cases can be assumed to be proportional to the Planck length \cite{AmelinoCamelia:2008qg}). Thus, in this context, an immediate question arises: may a MDR be a manifestation of a departure of the Riemannian\footnote{For simplicity, from now on we remove the word ``pseudo'' when referring to pseudo-Riemannian spaces.} nature of spacetime at a scale sensible for quantum gravity? If spacetime was non-Riemannian in a sense that could afford for a non-quadratic norm of vectors, then we may have a MDR, but the inverse is not necessarily true, therefore we can only speculate about it.
\par
A known branch of differential geometry that fits perfectly into this property is the so called \textit{Finsler geometry} \cite{HRund}. Basically, it is a rigorous formalization for possible non-quadratic norms of vectors, which is achieved through a four-velocity-dependent metric, but still preserving the parametrization invariance of the arc-length of curves in the manifold through the requirement of having a metric $g(x,\dot{x})$, homogeneous in the four-velocities. The relations between Finsler geometry and some approaches to the Quantum Gravity problem have been widely investigated \cite{Mavromatos:2007xe}, with special interest in models introducing some modification to particles' dispersion relation (see for instance Ref.\cite{Vacaru:2010fa} and references therein). In particular, in \cite{Girelli:2006fw} Girelli, Liberati and Sindoni pointed out that phenomenological models implementing some kind of departure from Lorentz symmetries can be systematically formalized within the framework of Finsler geometry, hypothesizing that such a formalism may be also the correct mathematical framework to describe the so called ``rainbow metrics" approach \cite{Magueijo:2002xx}.
\par
It is still unclear whether the introduction of momentum-dependent ``Rainbow" metrics imposes some sort of breakdown of Lorentz symmetry or otherwise they may be suitable for a scenario with a deformation of (local) spacetime symmetries through a deformed Poincar\'{e} group as in the Deformed Special Relativity approach \cite{AmelinoCamelia:2000mn}. This issue was preliminarily treated by the authors of Ref.\cite{Girelli:2006fw} for what concerns the fate of spacetime symmetries in MDR-inspired Finsler geometries and was explicitly considered in Ref.\cite{Amelino-Camelia:2014rga}, in which it was shown that it is possible to obtain a description of modified relativistic particle kinematics satisfying both Finsler geometry and Deformed Special Relativity prescriptions within the so called $\kappa$-Poincar\'{e} framework \cite{Lukierski1,Lukierski2}, at least at first order in the deformation parameter $\ell\sim 1/M_P$, where (in units such as $c=\hbar=1$) we can expect $M_P$ to be of the order of Planck scale $\sim 1.2 \cdotp 10^{28} eV$. It is worth mentioning that we are focusing on a first order deformation in $\ell$ for two different reasons. For a general theoretical case, in which one might consider exact Modified Dispersion Relations, the same analysis would need one to be able to write each conjugate momenta in terms of the four-velocities in order to perform the Legendre transform, which in general, at all orders in $\ell$, may not be possible. Moreover, given the small magnitude of Planck-scale effects on physical observables it is in general hard to outline some sort of phenomenology at order $\sim\ell^2$ (also first order effects clearly need some magnification mechanism in order to define falsifiable predictions). Therefore, in order to keep our approach generic and to describe a few simple phenomenological features it is sufficient to take into account deformations at the leading order of the deformation parameter.
\par
However satisfactory for what concerns the description of particles kinematics in the flat-spacetime limit, Ref.\cite{Amelino-Camelia:2014rga} leaves open the question whether this approach can be generalized to describe Deformed Relativity particles phenomenology \cite{AmelinoCamelia:1999zc,Jacobson:2002ye,AmelinoCamelia:2011bq,Amelino-Camelia:2016fuh} within a curved background. We would like here to implement this aspect, using as guidance some previous approaches \cite{Jack} to $\ell$-deformed particles kinematics in a totally-symmetric curved spacetime, where we will denote the parameter of spacetime curvature as $H$.
\par
A clear aspect presented in \cite{Girelli:2006fw} is the lack of a Finsler metric that would be probed through spacetime inferences with massless particles, because the massless limit of the presented metric is ill-behaved. For example, this behavior prevents us from having a metric probed by neutrinos, since, even being massive, for most practical reasons they are considered massless due to their tiny rest mass. In this paper, besides generalizing some results of Ref. \cite{Girelli:2006fw,Amelino-Camelia:2014rga}, we propose a different way of calculating a four-velocity-dependent (and still partially homogeneous in the velocities) metric, that can reproduce the main features of the standard Finsler case, like the dispersion relation, the geodesics as worldlines, the presence of DSR symmetries and, furthermore, presents a well-behaved massless limit, thus allowing for a generalized Finsler-like metric for any type of particle. Such generalization will allow us to propose an expression that deepens the one of Jacob and Piran \cite{Jacob:2008bw} for the MDR-induced time delay for particles with different energies, and to analyze the Finsler nature of spacetime using photons/neutrinos as probes. Furthermore, we will study in the last section the law of interaction between elementary particles compatible with DSR models, furnishing a description in terms of the tangent space, thus being compatible with the Finsler formalism.
%%%%%%%%%%%%%%%%%%%%%%%%%%%%%%%%%%%%%%%%%%%%%%%%%%%%%%%%%%%%%%

\subsection{Notations and Hamiltonian operator}

In this paper, in order to connect with the previous literature \cite{Jack,Amelino-Camelia:2014rga} and express the mathematical and physical concepts in an effective and comprehensible way, we will need to rely on different notations and coordinate sets.\\
The first one is the comoving Relative-Locality coordinate set for momentum-space $p_\alpha$ and spacetime flat slicing coordinates $x^\beta$, which satisfy the relation
\begin{equation}
\{p_\alpha , x^\beta\}=\delta_\alpha^\beta\,.
\end{equation}
In this coordinate set the Hamiltonian which formalizes the physics of particles embedded in a de Sitter-like curved spacetime, with $H$ the parameter of curvature and $\ell$ the deformation parameter due to Planck-scale effects, can be written as
\begin{equation}
\mathcal{H}=p_0^2-p_1^2 e^{-2 H x^0}+\ell\left(\gamma p_0^3+\beta p_0 p_1^2 e^{-2 H x^0}\right)\, ,\label{FlatHamilt}
\end{equation}
the generic MDR obtained imposing the on-shell relation to the Hamiltonian in (\ref{FlatHamilt}) contains the most general deformations, being $\ell$ the only scale available for deforming the mass-shell relation, that one can consider without having any implications on spacial rotations (no third power deformations of momentum $p_1$). This kind of Hamiltonian have been studied in many phenomenological explorations (mostly in conformal coordinates) such the ones in Ref \cite{Jack}.
\par
We stress that for sake of simplicity we are working in $1+1$ dimensions at first order in $\ell$ (since the same kinematical results are preserved in a higher number dimensions). In order to describe physics in an intuitive way, we will also consider the so called {\it conformal time coordinatization} expressed by coordinates $(\Omega,\Pi)$ and $(\eta,x)$. The relation between the previous coordinate set and this latter one is
\begin{eqnarray}
\eta&=&\frac{1-e^{-H x^0}}{H}\;,\;\;\;\;\;\;\; x=x^1\;,\\
\Omega&=& p_0 e^{H x^0}\;,\;\;\;\;\;\;\;\;\;\;\;\;\, \Pi=p_1\,.
\end{eqnarray}
One can verify the effectiveness of expressing physics in conformal-time coordinates by noticing that the spacetime line-element in this coordinatization can be expressed simply as $ds_\eta^2=(1-H\eta)^{-2}ds^2_{flat\, s-t}$.
Previous works analysing such deformed physical frameworks (see Ref.\cite{Jack}), express this same Hamiltonian (\ref{FlatHamilt}) depending on canonical variables $(q^{\mu},P_{\nu})$, deformation parameters $\gamma$ and $\beta$, deformation parameter $\ell $ and cosmological constant $H$. We define the energy and momentum respectively as $P_{\mu}=(\Omega, \Pi)$ and spacetime coordinates as $q^{\mu}=(\eta,x)$. Therefore, in conformal coordinates we have
\begin{equation}
{\cal H}_\Omega= (1-H\eta)^2(\Omega^2-\Pi^2 )+\ell(1-H\eta)^3\left(\gamma \Omega^3+\beta \Omega \Pi^2\right)\,.\label{hamiltonian}
\end{equation}
It is also easy to verify that $\Omega$ and $\eta$ (as well as $\Pi$ and $x$) still are conjugate variables.\\
The last couple of variables which will have some importance along this article is the one composed by the so called natural momenta, {\it id est} those that have the same functional dependence on local momenta as the charges of the translation generators of our spacetime:
\begin{eqnarray}
E&=&p_0-H x^1 p_1=\Omega(1-H\eta)-H x\Pi\,,\\
p&=&p_1=\Pi\,,
\end{eqnarray}
such charges can be easily obtained solving the de Sitter Killing equations, see for more explanations Refs.\cite{Marciano:2010gq,Jack,Amelino-Camelia:2013uya}. Using those variables the expression of the Hamiltonian (\ref{FlatHamilt}) becomes:
\begin{eqnarray}
{\cal H}_E &=& E^2-p^2+2 H p {\cal N}+\ell \beta E p\left(p-2 H{\cal N} \right)+\ell \gamma E\left(E^2+H p {\cal N}\right)\nonumber\\
&=& E^2-p^2+2 H p {\cal N}\left(1-\ell \left(\beta-\frac{\gamma}{2}\right)E \right)+\ell\left(\beta E p^2+\gamma E^3\right)\,.\label{HamTrasl}
\end{eqnarray}
The physics of particles described by the deformed Hamiltonians defined in this section was widely explored in Ref.\cite{Jack}, however many aspects of the formalization of those phenomena are missing in such approaches to the study of deformed relativistic frameworks, such as metric formalism, Killing vectors and particle interactions. We would like here to  provide a first exploration on how the integration of elements borrowed by the so called deformed momentum-space framework and Finsler geometry could enrich this kind of approaches to the study of deformed relativity theories, providing solid theoretical foundations to the formalism and also new suggestions on the phenomenological side.

%%%%%%%%%%%%%%%%%%%%%%%%%%%%%%%%%%%%%%%%%%%%%%%%%%%%%%%%%%%%%%

\section{Introduction on deformed Lagrangian formalism and symmetries}\label{sec:2}

Finsler geometry has  been introduced in previous articles \cite{Girelli:2006fw,Amelino-Camelia:2014rga} to formalize deformed relativistic frameworks defining a so called Finsler norm, a homogenous function ${\cal F}(\dot{x})$ on tangent space. In order to define a Finsler function, we can start writing down the particle's action explicitly in terms of the Hamiltonian
\begin{equation}
{\cal S}=\int \dot{x}^\alpha p_\alpha -\lambda\left({\cal H}-m^2\right) d\tau\,,\label{genAction}
\end{equation} 
where $\mathcal{H}$ is of course the Hamiltonian (\ref{FlatHamilt}), and $\lambda$ some Lagrange multiplier. In general a semi-classical metric structure is effectively encoded in Finsler geometry. On the other hand it turns out that $\kappa$-Poincar\'{e} group (that we are here using as foundation of a Planck-scale deformed particle dynamics in a maximally symmetric spacetime) does describe the symmetries of particles living on a de Sitter curved momentum-space \cite{AmelinoCamelia:2010qv,AmelinoCamelia:2011bm,AmelinoCamelia:2011cv,AmelinoCamelia:2011nt,Amelino-Camelia:2013uya,Loret:2014uia,Gubitosi:2013rna} and flat spacetime. The form of the momentum-space metric is chosen by imposing the invariant Hamiltonian to be the integration of the momentum space line element, this procedure can be easily generalized also at all orders in $H$ (see appendix for more details on momentum-space dynamics). 
 %\textcolor{teal}{We will here generalize the approach of \cite{Amelino-Camelia:2014rga} at all orders in the Hubble parameter $H$ to find the underlying Lagrangian formalism from which we will be able to define the Finsler norm for a Panck-scale deformed maximally symmetric curved spacetime}. 
 Those two different metric formalisms Finsler metric, $g^F_{\mu\nu}(\dot{x})$ and momentum-space metric, 

\begin{equation} 
 \zeta^{\alpha\beta}(p)=\left(\begin{array}{cc}
 1+2 \gamma\ell p_0  & 0\\
0 & -e^{-2 H x^0}(1-2\beta\ell p_0)
 \end{array}\right)\,
\end{equation}

could lead to confusion on the metric structures we refer to when we will introduce concepts such as geodesics and invariant line-element. Therefore we may need here to briefly review deformed-relativistic Lagrangian formalism and the role that those different metrics play in it, in the massless case for sake of simplicity.\\
The minimization of (\ref{genAction}) provide the relation between momentum and four-velocity (here in 1+1D):
\begin{equation}
\dot{x}^\alpha=\lambda(\dot{x})\{{\cal H},x^\alpha\}\,,\label{Hameq}
\end{equation} 
then at first order in $\ell$ is not hard to invert this relation, finding some $p(\dot{x})$, which makes possible to re-express the Casimir in terms of four-velocities:
\begin{equation}
{\cal H}(\dot{x})=\zeta_{\alpha\beta}(\dot{x})\dot{x}^\alpha\dot{x}^\beta\,,\label{Cxpunto}
\end{equation}
where $\zeta_{\mu\nu}(\dot{x})$ is the inverse of momentum-space metric expressed in terms of $\dot{x}$. It is also possible now to find an explicit solution for $\lambda(\dot{x})$, by imposing $\partial\mathcal{L}/\partial \lambda =0$. In the massive case this leads to
\begin{equation}
\lambda(\dot{x})=\frac{1}{2}\frac{\sqrt{\zeta_{\mu\nu}\dot{x}^\mu\dot{x}^\nu}}{m}\,.\label{lambxpunto}
\end{equation}
In the massless case, in general, it is not possible to solve $\partial\mathcal{L}/\partial \lambda =0$ with respect to $\lambda$, however taking into account (\ref{Cxpunto}) and (\ref{lambxpunto}) we can find very naturally that in the limit $m\rightarrow 0$, $\lambda\rightarrow 1/2$. In this limit then (\ref{Hameq}) in 1+1D become

\begin{equation}
\left\{\begin{array}{l}
\dot{x}^0=p_0+\frac{\ell}{2}\left(\beta p_1^2 e^{-2 H x^0}+3\gamma p_0^2\right)\\
\dot{x}^1=-p_1 e^{-2 H x^0}(1-\beta\ell p_0)
\end{array}\right.\,,\label{uxoux1}
\end{equation}  

and their inversion gives 

\begin{equation}
\left\{\begin{array}{l}
p_0=\dot{x}^0-\frac{\ell}{2}(\beta (\dot{x}^1)^2 e^{2 H x^0}+3\gamma (\dot{x}^0)^2)\\
p_1=-\dot{x}^1 e^{2 H x^0}(1+\beta\ell \dot{x}^0)
\end{array}\right.\,.
\end{equation}

While this inversion operation is straightforward at first order in $\ell$, in a non-perturbative scenario (at all orders in $\ell$) the explicit expression of $\dot{x}(p)$ may involve non invertible functions, which would make very hard (and in some cases impossible) to find the exact form of $p(\dot{x})$.
\par
Using those last results we find that the Lagrangian of our theory can be expressed as
\begin{equation}
\mathcal{L}=\dot{x}^\alpha p_\alpha -\frac{1}{2}\zeta_{\mu\nu}\dot{x}^\mu\dot{x}^\nu=\frac{1}{2}g_{\sigma\rho}(\dot{x})\dot{x}^\sigma\dot{x}^\rho\,,\label{genLag}
\end{equation}
where the Lagrangian metric $g$ is not univocally determined, as we will better discuss later, and should be chosen according to the requests of the theory. As we can see from (\ref{genLag}) in this kind of deformed-symmetry theories we loose the uniqueness of the metric for spacetime and momentum-space. Those different metrics play a different role in the theory and one should be always careful to use the right one in the right formula. For instance for Euler-Lagrange derived relations, such as geodesic equations or Killing equations we are referring to metric $g$. The light cone structure is instead characterized in terms of momentum-space metric, since the on-shell relation is $\zeta_{\mu\nu}\dot{x}^\mu\dot{x}^\nu=m^2$. Intriguingly, this defines the invariant (see Refs\cite{Amelino-Camelia:2014rga,Loret:2014uia}) element for flat spacetime as $\Delta s^2=\zeta_{\mu\nu}\Delta x^\mu \Delta x^\nu$.\\
It may be disappointing to have lost the possibility to express our theory with a unique metric, however from (\ref{genLag}) we can still identify a couple of interesting relations:
\begin{eqnarray}
\zeta_{\mu\nu}(\dot{x})\dot{x}^\mu\dot{x}^\nu &=& g^{\alpha\beta}(p)p_\alpha p_\beta\,,\\
\zeta^{\alpha\beta}(p)p_\alpha p_\beta &=& g_{\mu\nu}(\dot{x})\dot{x}^\mu\dot{x}^\nu\,,
\end{eqnarray}  
which highlight the duality between spacetime and momentum-space of this theory, widely discussed in the literature (see for instance \cite{Amelino-Camelia:2013uya}).\\
In order to fully understand whether deformed relativistic frameworks and Finsler geometry are physically equivalent, in \cite{Amelino-Camelia:2014rga} the symmetry transformations derived within the Finsler framework were identified with the ones generated by a deformed-Poincar\'{e} group. We want to show here how those generators can be obtained \textit{a priori} using Finsler Killing equations with metric $g$. 
%For the sake of simplicity, we will here find the boost $\mathcal{N}$ representation in the limit $H\rightarrow 0$ and then generalize this result to the case at all orders in $H$. 
Hamiltonian (\ref{FlatHamilt}) at order zero in $H$ is of course invariant under translations since
\begin{equation}
\{p_0,{\cal H}\}=0\;,\;\;\;\{p_1,{\cal H}\}=0\,,
\end{equation}
however if, always at zeroth order in $H$, we try to find the representation of the boost imposing
\begin{equation}
\{{\cal N},{\cal H}\}=0\,,
\end{equation}
we obtain two possible solutions:
\begin{eqnarray}
{\cal N}^{(1)}&=&x^0 p_1+x^1\left(p_0+\left(\beta+\frac{\gamma}{2}\right)\ell p_0^2+\beta\frac{\ell}{2}p_1^2\right)\,,\\
{\cal N}^{(2)}&=&x^0 p_1\left(1-\frac{\gamma}{2}\ell p_0\right)+x^1\left(p_0+\beta\ell p_0^2+\beta\frac{\ell}{2}p_1^2\right)\,,
\end{eqnarray}
or maybe some combination of the two. The situation at all orders in $H$ is of course even more complicated, which representation for the boost generator should one choose? A solution to this problem can be provided by the powerful Finsler formalism already explored in  the formalization of deformed relativistic frameworks in Ref.\cite{Girelli:2006fw,Amelino-Camelia:2014rga}. In Finsler geometry it is, in fact, possible to define a deformed Killing equation:

\begin{equation}
 \partial_\alpha g_{\mu\nu}\xi^\alpha + g_{\alpha\nu}\partial_\mu \xi^\alpha +  g_{\mu\alpha}\partial_\nu \xi^\alpha + \frac{\partial g_{\mu\nu}}{\partial\dot{x}^\beta}\partial_\alpha\xi^\beta \dot{x}^\alpha  =0\,.\label{KillingEq}
 \end{equation}
 
One can try to solve those equations for a generic four-velocity-dependent test metric such as 

\begin{equation}\label{beta_metric}
g_{\mu\nu}=\left(\begin{array}{cc}
1-\gamma\ell\dot{x}^0 & -e^{2 H x^0}\beta_2\ell\dot{x}^1\\
-e^{2 H x^0}\beta_2\ell\dot{x}^1 & -e^{2 H x^0}(1+\beta_1\ell\dot{x}^0)
\end{array}\right)
\end{equation}
Using \eqref{KillingEq}, after a few complicated steps the boost generator can be identified (see appendix for an example on how to solve the Finsler Killing equation in the flat spacetime case) as
\begin{eqnarray}
\mathcal{N}&=&\left[\frac{1-e^{-2 H x^0}}{2H}\left(1-\gamma\ell(p_0-H x^1 p_1)\right)-\frac{H}{2}(x^1)^2\left[1+2\left(\beta_1+2\beta_2\right)\ell p_0 \right]\right]p_1 + \nonumber\\
 &+&x^1\left[p_0+ \left(\beta_1+2\beta_2+\frac{\gamma}{2}\right)\ell p_0^2+\frac{\ell}{2}\left(\beta_1+2\beta_2\right)p_1^2 e^{-2 H x^0}\right] - \frac{\ell H}{2}\left(\beta_1+2\beta_2\right)(p_0-H x^1 p_1)p_1 (x^1)^2\,.
\end{eqnarray}
which in the flat spacetime limit, $H\rightarrow 0$, reduces to
\begin{equation}
\mathcal{N}_\ell=x^0 p_1\left(1-\ell\gamma p_0\right) + x^1\left(p_0+(\beta_1+2\beta_2+\frac{\gamma}{2})\ell p_0^2+\frac{\ell}{2}\left(\beta_1+2\beta_2\right) p_1^2\right)\,,\label{BoostRepresentation}
\end{equation}
in which we recognize a combination of the two candidates ${\cal N}^{(1)},{\cal N}^{(2)}$, while in the classical limit $\ell\rightarrow 0$ gives back the de Sitter boost generator (see for instance \cite{Marciano:2010gq}):
\begin{equation}
{\cal N}_H=\left(\frac{1-e^{-2 H x^0}}{2H}-\frac{H}{2}(x^1)^2\right)p_1+x^1 p_0\,.
\end{equation}
The Hamiltonian satisfying $\{{\cal N},{\cal H}\}$ can be obtained though the contraction of momenta with the metric $\tilde{g}(p)$:
\begin{equation}
\mathcal{H}=g^{\alpha\beta}(p)p_\alpha p_\beta=p_0^2-p_1^2+\ell\left((\beta_1+2\beta_2)e^{-2 H x^0} p_0 p_1^2 + \gamma p_0^3\right)\,.\label{genCasimir12}
\end{equation} 

In equation (\ref{genCasimir12}) we can recognize Hamiltonian (\ref{FlatHamilt}), under the condition $\beta=\beta_1+2\beta_2$, which introduces a freedom that we will have to constrain based on the requests of the theory. In general an approach based on curved momentum-spaces work perfectly fine with a diagonal metric, {\it id est} $\beta_1=\beta,\,\beta_2=0$ as well as with other choices. On the other hand, in order to formalize our theory (\ref{hamiltonian}) as a Finsler geometry we will need a homogeneous-like metric, which will force us to the choice $\beta_1=\beta_2=\beta/3$, as we will discuss further in this article.
The deformed de Sitter algebra of spacetime symmetry generators $E, p, \mathcal{N}$ at all orders in $H$ is characterized by the following Poisson brackets:
\begin{eqnarray}
&\{E,p\}=Hp\;,\;\;\;\{{\cal N},E\}=-p+H{\cal N}+\gamma\ell p E\,&\\
&\{{\cal N},p\}=-E-\ell\left(\frac{\beta}{2}p^2+\left(\beta+\frac{\gamma}{2}\right)E^2-H\left(\beta-\gamma\right){\cal N}p\right)\,.&
\end{eqnarray}
Of course those relations are, by definition, coherent with the Poisson brackets between generators and Hamiltonian (\ref{HamTrasl}):
$$
\{E,{\cal H}_E\}=\{p,{\cal H}_E\}=\{{\cal N},{\cal H}_E\}=0\,.
$$
It should be noticed in the end that the algebra described in this section is compatible with the well-known q-de Sitter algebra \cite{qdSpreliminary,Marciano:2010gq,Barcaroli:2015eqe}, at first order in $\ell$, under the choice $\beta= -1\,,\gamma=0$.
%\begin{eqnarray}
%\{p_0,p_1\}=0\;,\;\;\;\{{\cal N},p_0\}=-p_1 + \ell\gamma p_0 p_1\;,\\
%\{{\cal N},p_0\}=-p_0-\ell\left(\left(\beta+\frac{\gamma}{2}\right) p_0^2+\frac{\beta}{2}p_1^2\right)\,.
%\end{eqnarray}

%%%%%%%%%%%%%%%%%%%%%%%%%%%%%%%%%%%%%%%%%%%%%%%%%%%%%%%%%%%%%%
%%%%%%%%%%%%%%%%%%%%%%%%%%%%%%%%%%%%%%%%%%%%%%%%%%%%%%%%%%%%%%
%%%%%%%%%%%%%%%%%%%%%%%%%%%%%%%%%%%%%%%%%%%%%%%%%%%%%%%%%%%%%%

\section{From Hamiltonian formalism to a Finsler geometry}

We started with a non-trivial local structure of spacetime that could be modeled by a curved momentum-space, which is responsible for the definition of a modified dispersion relation and a deformed picture of interactions (that will be discussed in section \ref{comp_rule}). Now, we can analyze how a spacetime can emerge from these considerations, {\it id est} how the local structure can interfere on the geometry of the effective manifold that these high-energetic particles probe. We will follow a procedure that resembles the one presented in Refs. \cite{Girelli:2006fw,Amelino-Camelia:2014rga}. In these references, the authors perform a Legendre transformation that links the action of a particle from the Hamiltonian to the Lagrangian formalism. Since in Special and General Relativity one identifies the action of a massive particle with the arc-length of its trajectory, the authors generalized this assumption to the case of MDRs, defining a non-quadratic norm of the particle's four-velocity vector. Which can be modeled by the well studied Finsler geometry, that is defined by a four-velocity-dependent norm. The Finsler norm satisfies the usual properties:
\begin{equation}
\left\{\begin{array}{l}
   \mathcal{F}(\dot{x})\neq 0\;\;\; \text{if}\;\;\; \dot{x}\neq 0\\
   \;\\
   \mathcal{F}(\epsilon\dot{x})=|\epsilon|\mathcal{F}(\dot{x}) \label{FinslerConditions}
\end{array}\right. 
\end{equation}
The second one is rather important, since it expresses the homogeneity property of the norm, which in general implies
\begin{equation}
\dot x^{\mu} \frac{\partial \mathcal{F}^{2}}{\partial \dot x^{\mu}}=2 \mathcal{F}^{2}\,. \label{EulerTheoremF2}
\end{equation}
In \cite{Girelli:2006fw, Amelino-Camelia:2014rga} it was formalized that a MDR theory can be well expressed in terms of Finsler geometry, identifying the Finsler norm from the action integral as
\begin{equation}
{\cal S}=m\int \mathcal{F}(\dot{x})  d\tau =m \int \sqrt{g^F_{\mu\nu}(\dot{x}) \dot{x}^\mu\dot{x}^\nu} \,,
\end{equation}
where $g^F_{\mu\nu}$ is of course the Finsler metric. As we will see, this metric depends on the four-velocity and mass of the particle, however, it will not have a smooth limit for massless particles. Therefore, we propose an intermediate procedure that allows to define an action that serves both for the massive and massless cases, thus allowing a well-defined, unique metric that may describe a generalized Finsler spacetime, that is probed by high-energetic particles.

%%%%%%%%%%%%%%%%%%%%%%%%%%%%%%%%%%%%%%%%%%%%%%%%%%%%%%%%%%%%%%

\subsection{Standard Finsler metric in de Sitter spacetime}

%\begin{equation}\label{hamiltonian}
%{\cal H}=(1-H\eta)^2(\Omega^2-\Pi^2)+\ell (1-H\eta)^3(\gamma \Omega^3+\beta \Omega \Pi^2)-m^2.
%\end{equation}
In order to generalize the procedure just described to fully symmetric curved spacetimes, our starting point is the particle action in conformal coordinates with modified dispersion relation ${\cal H}_\Omega =m^2$,
\begin{equation}\label{ham_action}
{\cal S}[q,p,\lambda]=\int d\tau [\dot{\eta}\Omega+\dot{x}\Pi-\lambda( {\cal H}_\Omega-m^2)].
\end{equation}
This is the simple generalization of covariant mechanics in de Sitter spacetime. The equations of motion are 
\begin{eqnarray}
\dot{\eta}-\lambda \ell(1-H\eta)^3(3\gamma \Omega^2+\beta \Pi^2)-2\lambda(1-H\eta)^2\Omega=0,\\
\dot{x}-2\lambda\ell \beta(\-H\eta)^3\Omega \Pi+2\lambda(1-H\eta)^2\Pi=0,\\
\dot{\Omega}-2\lambda H(1-H\eta)(\Omega^2-\Pi^2)-3\lambda \ell H(1-H\eta)^2(\gamma \Omega^3+\beta \Omega \Pi^2)=0,\\
\dot{\Pi}=0.\label{p1}
\end{eqnarray}
Following the same procedures of \cite{Amelino-Camelia:2014rga}, {\it id est} performing the Legendre transformation, substituting $ p\rightarrow \dot{q}$, the Lagrangian is now
\begin{equation}
{\cal L}(q,\dot{q},\lambda)=\frac{\dot{\eta}^2-\dot{x}^2}{4\lambda(1-H\eta)^2}-\frac{\ell}{8\lambda^2}\frac{\beta\dot{\eta}\dot{x}^2+\gamma\,\dot{\eta}^3}{(1-H\eta)^3}+\lambda m^2.\label{lagr1}
\end{equation}
As we already commented in Sec. \ref{sec:2}, in order to carry out this operation one needs the explicit expression of $p(\dot{q})$, which in general cannot be obtained at all orders in $\ell$. On the other hand, the linearization guarantees that this is possible for any deformation of the Hamiltonian, that is why in the previous literature \cite{Girelli:2006fw,Amelino-Camelia:2014rga} the procedure linking the MDR with the Finsler norm has always been performed at the leading order.
Minimizing this Lagrangian with respect to $\lambda$ we get
\begin{equation}\label{lambda}
\lambda=\frac{\sqrt{\dot{\eta}^2-\dot{x}^2}}{2m(1-H\eta)}-\ell\frac{\beta\,\dot{\eta}\dot{x}^2+\gamma\,\dot{\eta}^3}{2(1-H\eta)(\dot{\eta}^2-\dot{x}^2)}
\end{equation}
and thus we can identify the Finsler norm related to Hamiltonian (\ref{hamiltonian}) as
\begin{equation}
{\cal F}(q,\dot{q})=\frac{\sqrt{\dot{\eta}^2-\dot{x}^2}}{1-H\eta}-\frac{m\ell}{2}\frac{\beta\,\dot{\eta}\dot{x}^2+\gamma\,\dot{\eta}^3}{(1-H\eta)(\dot{\eta}^2-\dot{x}^2)}.
\end{equation}
At this point we can make contact with geometry, identifying the semi-classical metric structure effectively encoded in the Finsler formalism. The Finsler metric is defined through a homogenous function on tangent space
\begin{equation}
g^F_{\mu \nu}=\frac{1}{2}\frac{\partial}{\partial \dot{q}^{\mu}}\frac{\partial}{\partial \dot{q}^{\nu}}{\cal F}^2(q,\dot{q}),\label{f-metric}
\end{equation}
so
\begin{eqnarray}
g^F_{00}=\frac{1}{(1-H\eta)^2}-\frac{m\ell}{2}\frac{3\beta\dot{\eta}(\dot{x})^4-\gamma(5\dot{\eta}^3\dot{x}^2-2\dot{\eta}^5-6\dot{\eta}\dot{x}^4)}{(1-H\eta)^2(\dot{\eta}^2-\dot{x}^2)^{5/2}},\\
g^F_{11}=-\frac{1}{(1-H\eta)^2}-\frac{m\ell}{2}\frac{\beta(2\dot{\eta}^5+\dot{\eta}^3\dot{x}^2)+\gamma(\dot{\eta}^5+2\dot{\eta}^3\dot{x}^2)}{(1-H\eta)^2(\dot{\eta}^2-\dot{x}^2)^{5/2}},\\
g^F_{01}=\frac{m\ell}{2}\dot{x}^3\frac{\beta(4\dot{\eta}^2-\dot{x}^2)+3\gamma \dot{\eta}^2}{(1-H\eta)^2(\dot{\eta}^2-\dot{x}^2)^{5/2}}.
\end{eqnarray}
It furnishes the de Sitter metric if $\ell =0$ and furnishes the metric of paper \cite{Amelino-Camelia:2014rga} for $H=0$, $\gamma=0$ and $\beta=-1$.

%%%%%%%%%%%%%%%%%%%%%%%%%%%%%%%%%%%%%%%%%%%%%%%%%%%%%%%%%%%%%%

\subsubsection{Geodesics}
The analysis done so far is valid for the case of massive particles, $m\neq 0$, since the limit $m\rightarrow 0$ corresponds to a singularity, implying that observables measured with such metric involving observations with particles with tiny mass, like neutrinos, would not be well defined.
\par
Despite this fact, it is possible to solve the geodesic equation for massless particles by making $m=0$ in the geodesic solution just like was done in Ref. \cite{Amelino-Camelia:2014rga}. In fact, now that we have a metric for a massive particle, we can minimize the arc-length to find its world line. This way, just as it is done in \cite{Amelino-Camelia:2014rga}, even though the massless limit does not exist for the definition of the metric, it is well-behaved for the expression of the geodesics. 
\par
In fact, one could solve the geodesic equation that arises by extremizing the action, which furnishes:

\begin{equation}
\ddot{q}^{\alpha}+\gamma^{\alpha}_{\mu\nu}\dot{q}^{\mu}\dot{q}^{\nu}+\frac{1}{2}g_F^{\mu\alpha}\frac{d}{dt}\left[\left(\frac{\partial}{\partial \dot{q}^{\mu}}g^F_{\gamma\beta}\right)\dot{q}^{\gamma}\dot{q}^{\beta}\right]=\frac{\dot{{\cal F}}}{{\cal F}}\left[\dot{q}^{\alpha}+\frac{1}{2}g_F^{\mu\alpha}\left(\frac{\partial}{\partial\dot{q}^{\mu}}g^F_{\gamma\beta}\right)\dot{q}^{\gamma}\dot{q}^{\beta}\right],
\end{equation}
for
\begin{equation}
\gamma^{\alpha}_{\mu\nu}=\frac{1}{2}g_F^{\alpha\beta}\left(g^F_{\mu\beta,\nu}+g^F_{\nu\beta,\mu}-g^F_{\mu\nu,\beta}\right).
\end{equation}
Using the known expressions from the Finsler geometry literature, that follow from Euler theorem for homogenous functions
\begin{equation}
\dot{q}^{\alpha}\frac{\partial g^F_{\mu\nu}}{\partial \dot{q}^{\alpha}}=\dot{q}^{\mu}\frac{\partial g^F_{\mu\nu}}{\partial \dot{q}^{\alpha}}=\dot{q}^{\nu}\frac{\partial g^F_{\mu\nu}}{\partial \dot{q}^{\alpha}}=0,
\end{equation} 
we can write the geodesic equation as
\begin{equation}
\ddot{q}^{\alpha}+\gamma^{\alpha}_{\mu\nu}\dot{q}^{\mu}\dot{q}^{\nu}=\frac{d}{dt}(\ln {\cal F})\dot{q}^{\alpha}.
\end{equation}
\par
As this Finsler function does not depend on the coordinate $x$, we conclude that $\Pi/m=\partial F/\partial \dot{x}$ is a first integral to this problem, this is equivalent to equation (\ref{p1}). As this formalism is invariant by reparametrization of the solution, we can choose the gauge $\eta(\tau)=\tau$. Doing it and substituting in the first constrain, we can solve the equation
\begin{align}
\Pi=-\frac{m\dot{x}}{(1-H \eta)\sqrt{\dot{\eta}^2-\dot{x}^2}}-\frac{m^2\ell (\beta+\gamma)\dot{\eta}^3\dot{x}}{(1-H \eta)\left(\dot{\eta}^2-\dot{x}^2\right)^2}=-\frac{m\dot{x}}{(1-H \tau)\sqrt{1-\dot{x}^2}}-\frac{m^2\ell (\beta+\gamma)\dot{x}}{(1-H \tau)\left(1-\dot{x}^2\right)^2}.
\end{align}
Integrating this equation with initial condition $x(0)=\bar{x}$ we obtain its simple solution in which incoming photons have signature $\Pi<0$:
\begin{eqnarray}\label{solmassive}
x(\eta)-\bar{x}&=&\frac{\sqrt{m^2+\Pi^2}-\sqrt{m^2+\Pi^2(1-H \eta)^2}}{H \Pi} + \ell (\beta+\gamma)\eta \Pi\left(1-\frac{H \eta}{2}\right)\,.
\end{eqnarray}
The same worldline with spatial origin in $\bar{x}$ can as well be obtained upon integration from $\bar{x}$ to $x(\eta)$, imposing the on-shell relation to the Finsler norm. For a massless, on-shell particle, $m=0$, $|\Pi|=\Omega+{\cal O}(\ell)$, the above equation is
\begin{eqnarray}\label{solmassless}
x(\eta)-\bar{x}&=&\eta-\ell (\beta+\gamma)\eta\Omega\left(1-\frac{H \eta}{2}\right).
\end{eqnarray}
This is exactly the worldline found in the literature from the Hamiltonian (\ref{hamiltonian}), see Ref. \cite{Jack}. As this is the straightforward generalization of Refs. \cite{Girelli:2006fw, Amelino-Camelia:2014rga}, it also possesses the same kind of formal difficulties with respect to the massless metric. However, it has been argued, for example in Ref. \cite{Assaniousssi:2014ota}, that physical observables calculated, of course, by a macroscopic observer, should be performed using the classical, Riemannian metric, and in this case, there would be no problem in having such ill-defined metric, because its effect (apart from its geodesic) would be unobservable. But we raise the possibility that the natural metric ``seen" by a particle, like the one we found with a Finslerian structure, should be the one considered for the calculation of observables in the new geometrical framework, since the act of expressing a metric probed by such particle should only make sense if there is an observer who probes such geometrical structure. Which allows us to a deeper conclusion that observations performed with particles with a non-trivial dispersion relation (no matter the origin of this deformation) should indicate a possible non-Riemannian structure of the spacetime where it propagates. Thereby, by generalizing the geometrical structure with which one describes a spacetime, the physical observables should change reciprocally.

%%%%%%%%%%%%%%%%%%%%%%%%%%%%%%%%%%%%%%%%%%%%%%%%%%%%%%%%%%%%%%

\subsection{Generalized Finsler-like metric}
In the previous section we performed a Legendre transformation on the action of a particle with a MDR until we found a version of the action that we could identify with the arc-length of the particle's trajectory in a certain geometry. Since this was a non-quadratic function, but still invariant under reparametrizations, we could identify a Finsler function and derive a metric from it. For the massive case, there was no problem in the definition of the metric, but for the massless one, the metric was ill-defined. Such behaviour is due to the fact that the arc-length is the Nambu-Goto-like version of the action and it is well-known that it can only properly describe and only makes sense its extremization for the massive case and since the metric that would be found should depend on the mass of the particle, it is natural that such kind of approach would be problematic for $m\rightarrow 0$ (which was not the case in Riemannian geometry.
\par
A possible solution for such impasse is the use of an unique action for both cases. Therefore, we propose to read a metric from the \textit{Polyakov-like action} of a point particle \cite{polchinski}. Such formalism was initially explored in Ref.\cite{Lobo:2016lxm}.
\par
In fact, by performing the Legendre transformation in (\ref{ham_action}), we ended up with the Lagrangian (\ref{lagr1}). As the Lagrangian is an analytic function, the action can be uniquely expressed by a Taylor expansion in the four-velocities
	\begin{equation}
	{\cal S}[q,\lambda] = \int d\tau\left[{\cal L}{\big |}_{\dot{q} = 0} + \frac{\partial \cal L}{\partial 	\dot{q}^{\mu}}{\bigg |}_{\dot{q} = 0}\dot{q}^{\mu} +\frac{1}{2!}\frac{\partial^2 \cal L}{\partial \dot{q}^{\mu}\partial \dot{q}^{\nu}}\bigg|_{\dot{q} = 0}\dot{q}^{\mu}\dot{q}^{\nu} +\frac{1}{3!}\frac{\partial ^3 \cal L}{\partial 	\dot{q}^{\mu}\partial\dot{q}^{\nu}\partial\dot{q}^{\gamma}}\bigg|_{\dot{q} = 0}\dot{q}^{\mu}\dot{q}^{\nu}\dot{q}^{\gamma} +...+\lambda m^2 \right],
	\end{equation}
where the zeroth and first order terms vanish as well as those of higher than the third order, since the Lagrangian (\ref{lagr1}) is a third-degree polynomial. Therefore the action can be expressed as 
	\begin{equation}
	{\cal S}[q,\lambda]=\int d\tau \left[\frac{1}{4\lambda}\, \tilde{g}_{\mu\nu}(q,\dot{q},\lambda)\, \dot{q}^{\mu} \dot{q}^{\nu} + \lambda 	m^2\right],
	\end{equation}
where
	\begin{equation} \label{metricg}
	\tilde{g}_{\mu\nu}(q,\dot{q}, \lambda) = g^0_{\mu\nu}(q) + g^1_{\mu\nu}(q, \dot{q}, \lambda),
	\end{equation}
and we have identified
	\begin{equation}
	\frac{1}{4\lambda} g^0_{\mu\nu}=\frac{1}{2!}\frac{\partial^2 \cal L}{\partial \dot{q}^{\mu}\partial \dot{q}^{\nu}}\bigg|_{\dot{q} = 0}, \; \frac{1}{4\lambda} g^1_{\mu\nu} = \frac{1}{3!}\frac{\partial ^3 \cal L}{\partial \dot{q}^{\mu} \partial \dot{q}^{\nu} \partial\dot{q}^{\gamma}}\bigg|_{\dot{q} = 0}\dot{q}^{\gamma}.
	\end{equation}
This defines a metric that depends on the point of the manifold and the four-velocity. As we will see, the field $\lambda$ allows to distiguish between the massive and massless cases.
\subsubsection{Massive case}
If $m\neq 0$, $\lambda$ is found to be (\ref{lambda}). If we substitute it into the action, we find
\begin{equation}
	{\cal S}[q]=m\int d\tau \sqrt{\tilde{g}_{\mu\nu}(q,\dot{q})\, \dot{q}^{\mu} \dot{q}^{\nu}},
\end{equation}
where the metric is
\begin{equation} \label{massiveg}
	\tilde{g}_{\mu\nu} =(1-H\eta)^{-2}
	\begin{pmatrix}
	1 - \ell\,  \gamma \frac{m\dot{\eta}}{\sqrt{\dot{\eta}^2-\dot{x}^2}} && -\frac{1}{3}\ell \beta\frac{m\dot{x}}{\sqrt{\dot{\eta}^2-\dot{x}^2}}\\
	-\frac{1}{3}\ell \beta\frac{m\dot{x}}{\sqrt{\dot{\eta}^2-\dot{x}^2}} &&  -1 - \frac{1}{3}\ell\, \beta\frac{m\dot{\eta}}{\sqrt{\dot{\eta}^2-\dot{x}^2}} \\
	\end{pmatrix}.
	\end{equation}
From this expression, one can already realize that such metric is well-defined in the massless limit, since the extra terms behave like $m/\sqrt{\dot{\eta}^2-\dot{x}^2}$,\footnote{It should be noticed that since the two terms go to zero with the  same velocity in the classical case, then at first order, in the massless limit $\ell m/\sqrt{\dot{\eta}^2-\dot{x}^2}\rightarrow \ell+{\cal O}(\ell^2)$ and the metric $\tilde{g}_{\mu\nu}$ has no singular behaviour.} whereas in the standard Finsler case we had $m/(\dot{\eta}^2-\dot{x}^2)^{5/2}$, which of course explodes in the massless regime.
\par
The extremization of this action furnishes the expected geodesics (which is necessary for consistency), given by eq. (\ref{solmassive}), which coincide with those obtained using Hamilton equations for the Hamiltonian \eqref{hamiltonian} subject to the mass-shell condition $\mathcal{H} = m^2$, in the same parametrization.

%%%%%%%%%%%%%%%%%%%%%%%%%%%%%%%%%%%%%%%%%%%%%%%%%%%%%%%%%%%%%%

\subsubsection{Massless case}

For $m=0$, the Lagrange multiplier $\lambda$ cannot be solved. However, we can absorb it into the definition of the parameter and define the usual {\it affine parameter} as $2\lambda d\tau=ds$. With this definition, the action assumes the familiar form
\begin{equation} \label{action-s}
	{\cal S}[q]=\frac{1}{2}\int ds\, \tilde{g}_{\mu\nu}(q,q')q'^{\mu}q'^{\nu},
\end{equation}
where $q'\doteq dq/ds$, generalizing the result obtained in (\ref{genLag}) to curved spacetimes, for the metric (\ref{beta_metric}) and $\beta_1=\beta_2=\beta/3$. Moreover, the extremization of this functional (taking into account the massless condition as a fundamental one, {\it id est} the on-shell condition ${\cal H}=0$ written in terms of the four-velocities $q'$) furnishes the geodesic equation (\ref{solmassless}).

%%%%%%%%%%%%%%%%%%%%%%%%%%%%%%%%%%%%%%%%%%%%%%%%%%%%%%%%%%%%%%

\subsubsection{Energy-momentum-dependent metric}

We have shown that a natural geometrical formalism that describes the spacetime probed by particles with a Modified Dispersion Relation consists in the use of a four-velocity-dependent metric, such that the Riemannian structure is recovered in the low-energy limit. To give another appearance to this approach we can use the definition of the conjugate momenta $P_{\mu}=\partial {\cal L}/\partial \dot{q}^{\mu}$ and substitute the four-velocities that appear in the metric by the energy-momentum of the particle. In fact, we end up with
\begin{equation}  \label{metricEM}
	\tilde{g}_{\mu\nu} = f^2(\eta)
	\begin{pmatrix}
	1-\ell \gamma\Omega/f(\eta) & \frac{1}{3}\ell\, \beta\Pi/f(\eta)\\
\frac{1}{3}\ell \beta\Pi/f(\eta) & -1-\frac{1}{3}\ell\, \beta\Omega/f(\eta) \\
	\end{pmatrix},
	\end{equation}
where $f(\eta)=(1-H\eta)^{-1}$. This is how our formalism takes the form of a ``Rainbow metric" \cite{Magueijo:2002xx}, where the metric that an observer assigns to the spacetime by using high-energetic particles depends on its energy-momentum. As can be seen, once the metric is written this way, it is evident the smooth limit between the massless and massive cases, since the dependence on the mass has disappeared, remaining just the momenta one. This was not the case for the same substitution in the standard Finsler case, as can be verified in Ref. \cite{Amelino-Camelia:2014rga} already in the flat case, which is still true by adding curvature to the spacetime.\footnote{Another realization of Rainbow gravity using vector dependent metrics can be found in \cite{Carvalho:2015omv} in the context of disformal transformations.}

%%%%%%%%%%%%%%%%%%%%%%%%%%%%%%%%%%%%%%%%%%%%%%%%%%%%%%%%%%%%%%
%%%%%%%%%%%%%%%%%%%%%%%%%%%%%%%%%%%%%%%%%%%%%%%%%%%%%%%%%%%%%%
%%%%%%%%%%%%%%%%%%%%%%%%%%%%%%%%%%%%%%%%%%%%%%%%%%%%%%%%%%%%%%

\section{Deformed Riemannian metric, time delay and redshift}
The Finsler geometry is a formulation in which the arc-length functional plays a central role, and from it, one can derive a metric tensor that depends homogeneously on tangent-space elements. In the previous Section, we have considered a way of defining a four-velocity-dependent metric that shares many properties of the Finsler one, but is derived from a Polyakov-like action for a particle that obeys a Modified Dispersion Relation and generates the same basic features, like the dispersion relation from a norm and geodesics. One of our achievements is the smooth limit that the metric presents when passing from the massive to the massless case.
\par

We would like to probe the effects of such metric structure observing the travel-time through cosmological distances of a particle sufficiently energetic. For the sake of measurements, since they are performed using a macroscopic apparatus, it is usually considered the classical Riemannian metric for the definition of observables \cite{Assaniousssi:2014ota}, for example, the proper time of an observer on Earth, which is used to derive the expression for the time-delay of high energetic photons emitted from a GRB \cite{Jacob:2008bw}. The formalism of Finsler geometry that we just described allows us to relax this hypothesis, as we will see in this section. %Since the metric is Riemannian, its components do not depend on the energy/frequency of the probed particle.
\par
The Finsler metric is a function that depends on spacetime points and vector fields, {\it id est} to calculate the inner product of two vectors at a given point, it is also necessary to determine a direction or a third vector at that point. Denoting $g^{v}$ the metric tensor that depends on $v\in \Gamma(T{\cal M})$, then the inner product of vectors $u$ and $w$ is $g^v_{\ \alpha\beta}u^{\alpha}w^{\beta}$, it is then, necessary to specify the three vectors $(v,u,w)$ to measure the inner product. \textit{However, for each fixed $v_{0}$, we have that $g^{v_0}$ is a Riemannian metric.} Therefore, it is possible to calculate a vector-directed Riemannian scalar product at a given point for a fixed $v_0$, if this vector is defined at that point. See for instance Ref.\cite{anisotropic} for an analysis of vector dependent tensor calculus.
\par
%As aforesaid, in formalisms with energy-dependent or ``Rainbow" metrics, this metric is considered as the one that a particle ``probes", in this formalism, however, the observer is formalized as a low energetic worldline, therefore embedded in a momentum-independent classical metric.
%but the one that an observer assigns to the spacetime needs to be the classic one, because if it also was rainbow-like, the observer's metric would depend on the product between the observer's mass and the Planck length, which would be in obvious contradiction with the current observations\footnote{Or it would depend on the product between the observer's mass and $\ell/N$, where $N$ is the number of elementary particles that constitute the observer, which gives a negligible contribution.}.
\par

The formalism here described allows us to define a Riemannian metric, induced by a Finsler one by fixing the vector which the metric depends on as the tangent vector of the integral curve of the particle that probes the ``Rainbow" metric, {\it id est} $\tilde{g}\doteq g^{v_0}$, where $v_0$ is the particle's four-velocity. It is the \textit{Riemannian metric constructed by an observer by means of measurements with a particle with four-velocity $v_0$.}
\par

To visualize the reasonability of this proposal, consider the following example. Suppose that the Planck scale was not so far from our experience, such that spacetime inferences could always be performed with particles that manifest modified dispersion relations. Since we are hypothesizing that the spacetime probed by such particles is deformed with respect to the standard Riemannian one by means of a momentum-dependent metric, if an observer Alice that makes inferences using these particles intends to preserve the equivalence principle, she should assign such metric structure to the spacetime manifold. For example, the geometrical locus of the photon sphere of a Black Hole (BH) is a property of the photon's worldlines, and if she wants to preserve the equivalence principle by representing them as geodesics of a spacetime, they would need to be compatible with the metric structure of the spacetime surrounding a BH, which would be of Rainbow nature. In other words, the map of a spacetime constructed by such observer would need to inform such energy-dependent behaviour.\footnote{The study of the role of the equivalence principle in Finsler geometry in theories with MDR is still at an early stage. An interested reader may find some insight in Ref.\cite{LiberatiLetizia}.}
\par
So, following this principle, another example consists in the measurement of the proper-time, elapsed on Earth from the emission of a particle from a cosmological source up to its arrival at a terrestrial detector. As the spacetime geometry inferred using a particle with the same energy is determined by the deformed metric, the proper-time is calculated as $\int\sqrt{\tilde{g}_{\mu\nu}\dot{q}^{\mu}\dot{q}^{\nu}}d\lambda$, which in the comoving frame is $\int\sqrt{\tilde{g}_{00}}d\eta$, where $\tilde{g}$ is the Rainbow metric that depends on the energy of the detected particle.\footnote{The effect of considering a deformed metric for this calculation can be equivalently cast by absorbing this contribution in a redefinition of the emission or detection parameter $\eta$ and considering the standard Riemannian metric.}
\par
Therefore, for the measurement of the time delay between massless particles from GRBs \cite{AmelinoCamelia:1999zc,Amelino-Camelia:2016fuh} (which is one of the standard sources of constrains for MDRs), we define it analogously to the Shapiro delay \cite{wald}. In this case, the delay was calculated as the difference in the proper time of Earth between the time taken for a radar signal to reach a target and return depending if there is a massive object that deforms the spacetime geometry probed by the photon and the observer, {\it id est} in the sending and receiving of the signal one compares the proper times calculated in first and zeroth order in the mass parameter, for example in a Schwarzschild spacetime. Equivalently, in this model, depending on the energy of the probed particle, the induced metric will be the ordinary de Sitter one induced by a standard dispersion relation or will manifest a first order correction if the particle is sufficiently energetic, induced by the corresponding MDR. In our case, the parameter is not the mass in the Schwarzschild metric, but it is the Planck length $\ell$. %which is similar to the first order correction in the mass that deforms the flat metric in the Shapiro delay. 

%%%%%%%%%%%%%%%%%%%%%%%%%%%%%%%%%%%%%%%%%%%%%%%%%%%%%%%%%%%%%%%%%%%%%%%%%%%%%%%%%%%%%%%%%%%%%%%%%%%%%%%%%%%%%%%%%%%%%%%%%%%%%%%%%%%%%%%%%%%%%%%%%%%%%%%%%%%%%%%%%%%%%%%%%%%%%%%%%%%%%%%%%%%%%%%%%%%%%%%%%%%%%%%%%%%%%%%%%%%%%%%%

\subsection{On the propagation time}
\par
What we define as a time delay for the arrival of photons is the difference between the proper times (measured on Earth) that record the time elapsed from the emission of massless particles at $A$, to their arrival at the detector, labeled as $B$. For each photon, the proper time elapsed is
\begin{equation}
\tau = \int_A^B\sqrt{\tilde{g}_{00}} d\eta.
\end{equation} 
In this case, $\tilde{g}_{00}$ is the metric's component in the deformed Riemannian geometry induced by a Finsler-like metric that the observer associates to the spacetime from eq. (\ref{metricEM}), where the photon's four-velocity/momenta is the fixed vector that induces the deformed Riemannian metric. In the coordinates that we have been using $\tilde{g}_{00}=(1-H\eta)^{-2}-\ell \gamma\Omega(1-H\eta)^{-1} $. Integrating the proper time,
\begin{equation}
\tau = \frac{1}{H}\ln\left|{\frac{1-H \eta_A }{1- H\eta_B }}\right|-\ell\frac{\Omega}{2}\gamma(\eta_B-\eta_A).
\end{equation}
For sake of simplicity we assume the classical expression for the redshift $z$
\begin{equation}
1+z=a(t_B)/a(t_A)=\frac{1-H \eta_A }{1- H \eta_B}\,,
\end{equation}
in order to  keep explicit the Planck scale corrections in the time-delay formulation. An approach assuming a non-trivial expression for $z$ can be found for instance in \cite{Barcaroli:2015eqe}.

%%%%%%%%%%%%%%%%%%%%%%%%%%%%%%%%%%%%%%%%%%%%%%%%%%%%%%%%%%%%%%%%%%%%%%%%%%%%%%%%%%%%%%%%%%%%%%%%%%%%%%%%%%%%%%%%%%%%%%%%%%%%%%%%%%%%%%%%%%%%%%%%%%%%%%%%%%%%%%%%%%%%%%%%%%%%%%%%%%%%%%%%%%%%%%%%%%%%%%%%%%%%%%%%%%%%%%%%

\subsection{Calculating the time delay for the perturbed metric}
\par
\begin{figure}[h!]
\includegraphics[scale=0.9]{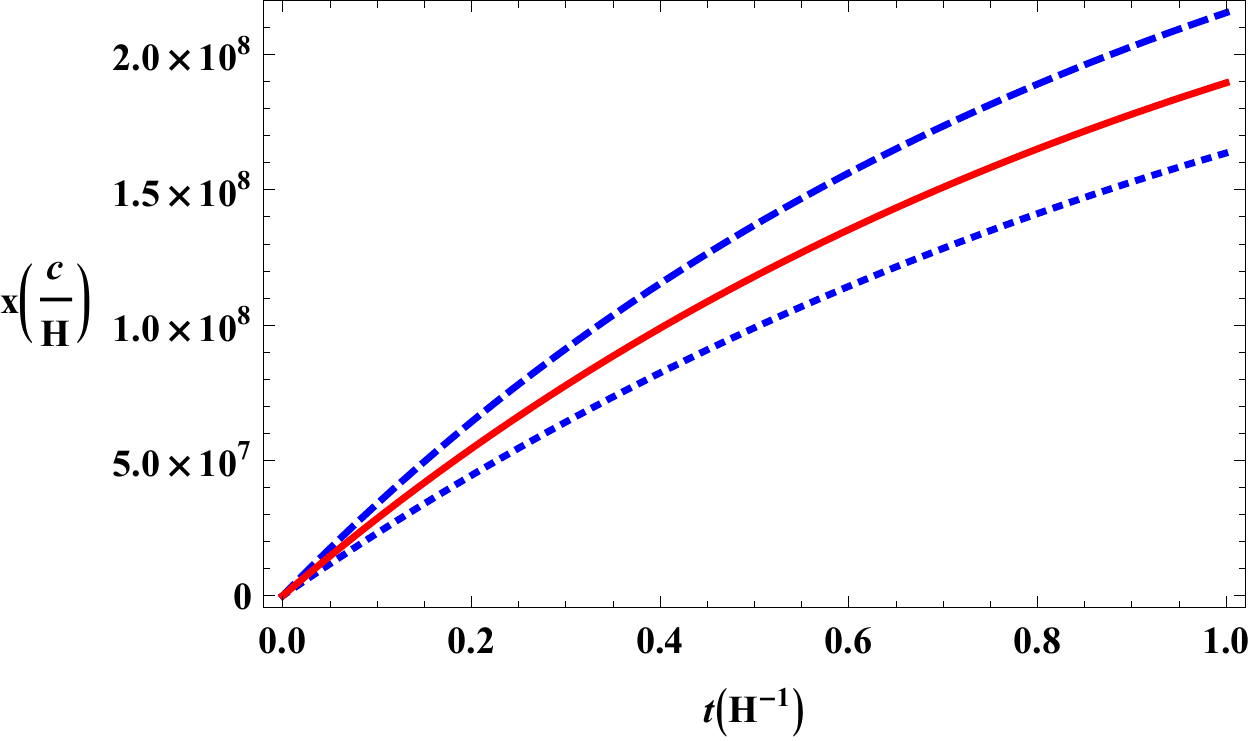}
\caption{\footnotesize Two massless particles, a high energetic one (blue lines) and a low energetic one (red line), are emitted simultaneously at the origin. We can observe that the high energetic particle anticipates (dashed line) or follows (dotted line) the infrared one, whether the sum of parameters $\beta$ and $\gamma$ is respectively $\beta+\gamma<0$ or $\beta+\gamma>0$.}
\label{fig:timedelay}
\end{figure}

Now, consider that we have a hard and a soft massless particle, for which $\Omega^s\ell \ll 1$, and $\Omega^h \ell \lesssim 1$, their worldlines will be
\begin{eqnarray}
x_h(\eta)&=&\eta-\ell (\beta+\gamma)(\eta-\eta_A)\Omega^h\left[1-\frac{H (\eta-\eta_A)}{2}\right],\\
x_s(\eta)&=&\eta.
\end{eqnarray}
Suppose that both particles are emitted simultaneously, {\it id est} $\eta_{sA}=\eta_{hA}\doteq -\eta_A$, at the same spatial point $x_s(\eta=\eta_A)=x_h(\eta=\eta_A)=x_A$ (see Figure \ref{fig:timedelay}). Let $B$ be the point where the detector on Earth is. The elapsed parameter necessary for the soft particle to reach the detector, which is located at the origin is $\eta_A$.

\par
To determine the elapsed parameter, denoted by $\eta_{hB}$, necessary for the hard trajectory to reach the same spacial point, we consider
\begin{eqnarray}
x_h(\eta_{hB})=0\Rightarrow \nonumber \\
\eta_{hB}=\ell (\beta+\gamma)\Omega^h\left(1+\frac{H \eta_{A}}{2}\right)\eta_{A}.\label{x0hb}
\end{eqnarray}
Since we normalize the scale-factor for $\eta_B=0$, and the redshift $z$ is an experimental parameter determined by soft particles, it is natural to define it as
\begin{equation}
1+z=1-H \eta_{sA}=1+H \eta_A\Rightarrow \eta_{A}=\frac{1}{H}z.\label{x0sb}
\end{equation}
So, the proper time elapsed on Earth until the arrival of the hard particle can be calculated using equations (\ref{x0hb}) and (\ref{x0sb}) as
\begin{eqnarray}
\tau_h&=&\frac{1}{H}\ln \left|\frac{1-H \eta_{hA }}{1- H \eta_{hB}} \right| -\ell\frac{\Omega^h}{2}\gamma(\eta_{hB}-\eta_{hA})= \frac{1}{H}\ln \left|\frac{1+H \eta_{A }}{1- H \eta_{hB}}\right|-\ell\frac{\Omega^h}{2}\gamma(\eta_{hB}+\eta_{A}) \Rightarrow\nonumber \\
\tau_h&\approx&\tau_s+\frac{\ell \Omega^h}{H}\left[(\beta+\gamma)(z+\frac{z^2}{2})-\frac{\gamma}{2} z\right].
\end{eqnarray}
The time delay is defined as the difference between the proper times $\tau_h$ and $\tau_s$, therefore it is
\begin{equation}
\Delta \tau\doteq \frac{\ell \Omega^h}{H}\left[(\beta+\gamma)(z+\frac{z^2}{2})-\frac{\gamma}{2} z\right].
\end{equation}
Such expression contains the usual one used for the calculation of time-delays presented in \cite{Jacob:2008bw}, added to an extra contribution due to the deformed metric that the observer assigns to spacetime when performing measurements with high-energetic particles. The previous approach could not distinguish between the parameters $\gamma$ and $\beta$, since the effect appeared as a factor of $\gamma+\beta$. Our new proposal allows to distinguish them, and introduces a phenomenological rule for the time-delay in a more flexible shape

\begin{equation}
\Delta \tau(z)= \frac{\ell \Omega^h}{H}(\xi_1\, z\, +\, \xi_2\, z^2),
\end{equation}
in our case, $\xi_1=\beta+\gamma/2$ and $\xi_2=(\gamma+\beta)/2$.
\par
In the de Sitter case, the use of Jacob and Piran's ansatz \cite{Jacob:2008bw} (that does not make reference to a four-velocity-dependent metric) predicts the time-delay
\begin{equation}
\Delta \tau_{J-P}(z)=\frac{\ell \Omega^h}{H}\xi\left(z+\frac{z^2}{2}\right),
\end{equation}
for $\xi=\beta+\gamma$.
%%%%%%%%%%%%%%%%%%%%%%%%%%%%%%%%%%%%%%%%%%%%%%%%%%%%%%%%%%%%%%%%%%%%%%%%%%%%%%%%%%%%%%%%%%%%%%%%%%%%%%%%%%%%%%%%%%%%%%%%%%%%%%%%%%%%%%%%%%%%%%%%%%%%%%%%%%%%%%%%%%%%%%%%%%%%%%%%%%%%%%%%%%%%%%%%%%%%%%%%%%%%%%%%%%%%%%%%

\subsection{Energy dependent redshift effect}
Another interesting correction is the redshift of the frequency of photons in this geometry. Different from the previous case in which the metric played a central role, for calculating this quantity it is only necessary to consider the equations of motion of the photon. In fact, in comoving coordinates\footnote{Now we use comoving coordinates because they are the standard ones used in the literature to measure this effect \cite{Amelino-Camelia:2013uya}.}, the equations of motion for spacetime coordinates and the energy for a massless particle are (\ref{uxoux1})
\begin{equation}\label{uxoux2}
\left\{\begin{array}{l}
\dot{x}^0=p_0+\frac{\ell}{2}\left(\beta p_1^2 e^{-2 H x^0}+3\gamma p_0^2\right)\\
\dot{x}^1=-p_1 e^{-2 H x^0}(1-\beta\ell p_0)\\
\dot{p}_0=-Hp_0^2(1+\ell\gamma p_0)
\end{array}\right.\,.
\end{equation}  
The solution for the energy is 
\begin{equation}\label{energy_massless}
p_0(\tau)=\frac{\bar{p}_0}{1+H\bar{p}_0\tau}-\ell\gamma\frac{\bar{p}_0^2}{(1+H\bar{p}_0\tau)^2}\ln(1+H\bar{p}_0\tau),
\end{equation}
where $\bar{p}_0$ is the energy of the photon for $\tau=0$. 

\begin{figure}[!h]
\includegraphics[scale=0.9]{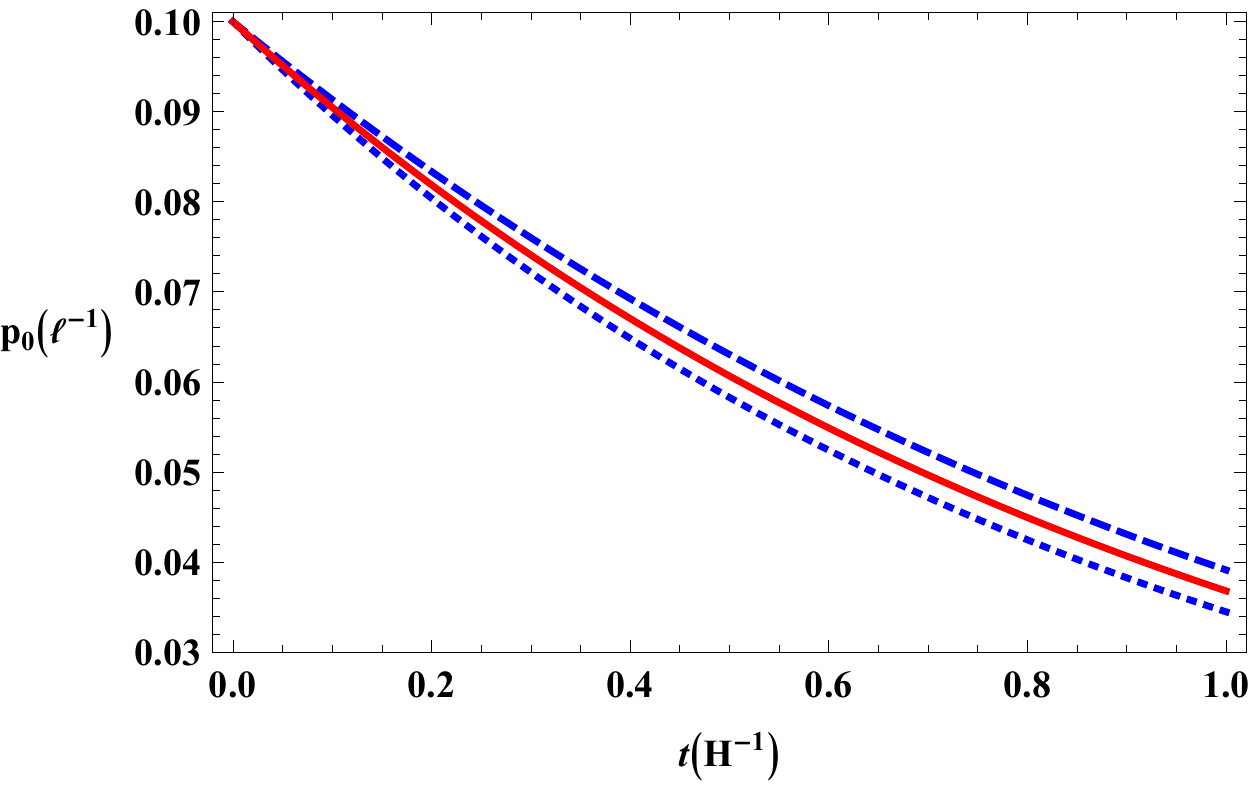}
\caption{\footnotesize Particles with different energies experience a different cosmological redshift. The dashed-blue line shows the deformation effect for $\beta+\gamma>0$, while the dotted-blue one for $\beta+\gamma<0$. The red line represents the classical undeformed cosmological redshift effect, for a low-energetic particle.}
\label{fig:lateshift}
\end{figure}
\par

For on-shell massless particles we have ${\cal H}=0$, which is simply 
\begin{equation}\label{on-shell-massless}
p_1^2=e^{2Hx^0}p_0^2\left[1+\ell(\beta+\gamma)p_0\right].
\end{equation} 
Substituting equation (\ref{energy_massless}) and (\ref{on-shell-massless} )in the first equation of (\ref{uxoux2}), we find
\begin{equation}
\Delta x^0(\tau)=\frac{1}{H}\ln(1+H\bar{p}_0\tau)+\ell\frac{\bar{p}_0}{1+H\bar{p}_0\tau}\left[\frac{1}{2}(\beta+\gamma)\bar{p}_0\tau+\frac{\gamma}{H}\ln(1+H\bar{p}_0\tau)\right],
\end{equation}
which can be solved for the parameter $\tau$ as
\begin{equation}
\tau= \frac{e^{H\Delta x^0}-1}{H\bar{p}_0}-\ell\left[(\beta+\gamma)\frac{e^{H\Delta x^0}-1}{2H}+\gamma\Delta x^0\right].
\end{equation}
Substituting this expression in the dependence of the energy with the parameter $\tau$ given by (\ref{energy_massless}), we derive the cosmological redshift of the frequency of a photon that obeys a modified dispersion relation (see Figure \ref{fig:lateshift})
\begin{equation}
p_0=\bar{p}_0e^{-H\Delta x^0}+\frac{\ell}{2}\bar{p}_0^2e^{-2H\Delta x^0}(\beta+\gamma)(e^{H\Delta x^0}-1).
\end{equation}

%%%%%%%%%%%%%%%%%%%%%%%%%%%%%%%%%%%%%%%%%%%%%%%%%%%%%%%%%%%%%%
%%%%%%%%%%%%%%%%%%%%%%%%%%%%%%%%%%%%%%%%%%%%%%%%%%%%%%%%%%%%%%
%%%%%%%%%%%%%%%%%%%%%%%%%%%%%%%%%%%%%%%%%%%%%%%%%%%%%%%%%%%%%%
  
\section{Nontrivial momentum composition rules}\label{comp_rule}

It was explained in \cite{Amelino-Camelia:2014rga}, using an argument already introduced in \cite{Gubitosi:2013rna}, that if one assumes deformed spacetime symmetries, then if we assume that whenever two particles interact, different observers must agree on the existence of the interaction vertex, this latter may be deformed too. This feature is formalized in the $\kappa$-Poincar\'{e} framework, introducing a deformed composition law for momenta
\begin{equation}
k_\mu=(p\oplus q)_\mu\, ,\label{nontrivcomp}
\end{equation}
which we can imagine as describing the decaying of a particle with momentum $k$ in two different particles with momenta $p$ and $q$. At first order in the deformation parameter $\ell$ we can express all the possible composition rules of momenta (requiring invariance under the parity transformation $p_0\rightarrow p_0$ , $p_1\rightarrow -p_1$, see \cite{Carmona:2012un}) relying just on four parameters $a,\, b,\, f\, g$ as
\begin{eqnarray}
(p\oplus q)_0&=&p_0+q_0+ \ell(a p_0 q_0 + b p_1 q_1)\,,\\
(p\oplus q)_1&=&p_1+q_1 + \ell(f p_0 q_1 + g p_1 q_0)\,.
\end{eqnarray}
Such deformed composition rules are known to be related to the coproduct of translation generators. Then, in order to determine which coalgebric sector can be compatible with our deformed boost generator (\ref{BoostRepresentation}), we need to find some condition which would allow us to find some relations between the composition rule parameters. Those relations can be found by requiring the composition of the boosted momenta to be invariant under the action of the boost
\begin{equation}
O'=O +\xi\{\mathcal{N},O\}\,.\label{Boostaction}
\end{equation}
However, before checking whether different (boosted) observers agree on the existence of the particles' vertex, we should pay attention to the possibility to have the so called {\it backreaction} effect \cite{Gubitosi:2013rna,Majid:2006xn,Carmona:2012un}, since the rapidity parameter $\xi$ can in general change in a momentum-dependent way, compatible with the coproducts of momenta and the action of Lorentz transformations on momenta themselves. This effect is defined as the right action $\vartriangleleft\,: \mathds{R} \times \Sigma\rightarrow\mathds{R}$, that we can parametrize as
\begin{equation}
\xi_1=\xi\vartriangleleft q =\xi(1+ \ell C q_0+\ell D q_1)\;,\;\;\;\xi_2=\xi\vartriangleleft p =\xi(1 + \ell B p_0+\ell A p_1)\,.\label{rightactions}
\end{equation}
Therefore, in general we will have
\begin{eqnarray}
&(p\oplus q)'_\mu(\xi)\neq p'(\xi)\oplus q'(\xi)\, ,&\\
&(p\oplus q)'_\mu(\xi) = p'(\xi\vartriangleleft q)\oplus q'(\xi\vartriangleleft p)\, .&\label{rightsumrule}
\end{eqnarray}
Now, imposing the complete parametrization of vertex transformation (\ref{rightsumrule}) with (\ref{Boostaction}) and (\ref{rightactions}),  we are able to find four relations between our eight parameters $a,b,f,g,A,B,C,D$, which ensure the existence od the vertex for any observer:
\begin{equation}
\left\{\begin{array}{ll}
A=0\\
D=0\\
a=-\gamma\\
b=\beta-B-C\\
f=\beta-C\\
g=\beta-B\\
\end{array}\right.\,.\label{rappparamgener}
\end{equation}
Those relations are perfectly compatible with the Hamiltonian invariance under boost:
\begin{equation}
{\cal H}(p\oplus q)={\cal H}(k)={\cal H}'(k)={\cal H}(k')={\cal H}(p'(\xi\vartriangleleft q)\oplus q'(\xi\vartriangleleft p))\, ,
\end{equation}
which shows {\it a posteriori} the coherence of those composition laws with the curved momentum-space framework. It can be noticed that our results \eqref{rappparamgener} are compatible with those at equations $(19)$ and $(20)$ of Ref. \cite{Carmona:2012un} in the $1+1$ dimensional case.\footnote{The equations $(19)$ and $(20)$ of Ref.\cite{Carmona:2012un} in terms of our parameters are in fact $a=-\gamma$, $\beta=f+g-b$ and $\gamma+\beta+a+b-f-g=0$, which are trivially satisfied by \eqref{rappparamgener}.} The problem is if also Finsler geometry can be compatible to such ``deformed sum". In Finsler geometry, in fact, the particle trajectory depends on spacetime coordinates $x^\alpha$ and four-velocities $\dot{x}^\beta$, which live in a flat tangent space to the curved spacetime. On the other hand, in  Relative Locality \cite{AmelinoCamelia:2011bm,AmelinoCamelia:2011nt,Amelino-Camelia:2013uya} momenta are the fundamental observables and spacetime is defined as the cotangent bundle to the curved momentum-space. If the deformation of momenta composition law fits very well with the curved momentum-space framework, one should not give for granted for this to happen also in the Finsler case. However in this section we showed that non trivial composition laws are imposed by the deformation of spacetime symmetries that arise from our theory. The impossibility to express this feature also for Finsler formalism would then jeopardise our entire argument.  

%%%%%%%%%%%%%%%%%%%%%%%%%%%%%%%%%%%%%%%%%%%%%%%%%%%%%%%%%%%%%%%%%%%%%%%%%%%%%%%%%%%%%

\subsection{Four-velocity tangent space}

From our perspective, the problem of the tangent space flatness is not a real problem. In fact every local observer is flat, and curvature arises only when confronting observations between different reference frames. Assuming then (\ref{nontrivcomp}), we are treating different particles as different reference frames in momentum-space, which can be characterized also as a non-trivial composition for four-velocities, formalizing the relations between the different locally-flat reference frames. This feature can be obtained taking into account a two-parameters family of coproducts and momenta compositions that we can obtain from (\ref{rappparamgener}):
\begin{eqnarray}
\Delta p_0&=&p_0\otimes \mathds{1} + \mathds{1}\otimes p_0-\ell(\gamma p_0\otimes p_0 - (\beta-B-C) p_1\otimes p_1)\,,\\
\Delta p_1&=&p_1\otimes \mathds{1}+\mathds{1}\otimes p_1 + \ell\left(\left(\beta-C\right) p_0\otimes p_1 + \left(\beta-B\right) p_1\otimes p_0\right)\,,\label{coprodotti2par}
\end{eqnarray}
and 
\begin{eqnarray}
(p\oplus q)_0&=&p_0+q_0- \ell(\gamma p_0 q_0 -(\beta-B-C) p_1 q_1)\,,\label{composition0}\\
(p\oplus q)_1&=&p_1+q_1 + \ell\left(\left(\beta-C\right) p_0 q_1 + \left(\beta-B\right) p_1 q_0\right)\,.\label{composition1}
\end{eqnarray}
Then, using the expression of four-velocities with respect to momenta (\ref{uxoux1}) with (\ref{composition0}) and (\ref{composition1}), in the massless case, we can easily obtain the composition laws for $\dot{x}$:
\begin{eqnarray}
\dot{x}^0(p)\oplus_{\dot{x}}\dot{x}^0(q)&=&\dot{x}^0(p\oplus q)=\dot{x}^0(p)+\dot{x}^0(q)+ 2\gamma\ell\dot{x}^0(p)\dot{x}^0(q) +\ell(2\beta-B-C)\dot{x}^1(p)\dot{x}^1(q)\,,\hspace{1.5cm}\label{comp1}\\
\dot{x}^1(p)\oplus_{\dot{x}}\dot{x}^1(q)&=&\dot{x}^1(p\oplus q)=\dot{x}^1(p)+\dot{x}^1(q) -\ell B\dot{x}^1(p)\dot{x}^0(q) -\ell C\dot{x}^0(p)\dot{x}^1(q).\label{comp2}
\end{eqnarray}
Those laws satisfy the relation
\begin{equation}
\dot{x}^\alpha((p\oplus q)'(\xi))=\dot{x}^\alpha(k')=\dot{x}^\alpha(p'(\xi\vartriangleleft q)\oplus q'(\xi\vartriangleleft p))= (\dot{x}^\alpha(p))'(\xi\vartriangleleft q)\oplus_{\dot{x}}(\dot{x}^\alpha(q))'(\xi\vartriangleleft p) =(\dot{x}^\alpha(p)\oplus_{\dot{x}}\dot{x}^\alpha(q))'(\xi)\,,
\end{equation}
where the backreaction has the exact same form of the one described in the previous section. A trivial consequence of the definition of velocity and the
composition rules \eqref{comp1} and \eqref{comp2}, that we can be found {\it a posteriori}, in the massless case is
\begin{equation}
\frac{\dot{x}^1(p)\oplus_{\dot{x}}\dot{x}^1(q)}{\dot{x}^0(p)\oplus_{\dot{x}}\dot{x}^0(q)}=1-(\beta+\gamma)\ell(p_0+q_0)\equiv v(p\oplus q)\, ,
\end{equation}
in which $v$ is the velocity that we already found in our massless worldlines (\ref{solmassless}). This structure of relations between different tangent spaces allows us to identify the phase-space invariant element as:
\begin{eqnarray}
\zeta_{\alpha\beta}(p\oplus q)\dot{x}^\alpha(p\oplus q)dx^\beta &=&\zeta_{\alpha\beta}(p'\oplus q')\dot{x}^\alpha(p'\oplus q')d(x^\beta)'=\zeta_{\alpha\beta}((p\oplus q)')\dot{x}^\alpha((p\oplus q)')d(x^\beta)'=\nonumber\\
&=&\zeta_{\alpha\beta}((p\oplus q)')((\dot{x}(p))'\oplus_{\dot{x}}(\dot{x}(q))')^\alpha d(x^\beta)'\,.
\end{eqnarray}
\par
The doubt we had at the beginning of this section was whether it is possible to express departures from Poincar\'{e} symmetries,  often formalized in the literature as a deformed Hamiltonian framework ($\ell$-deformed phase space), within the context of the Lagrange-Finsler geometry (tangent space). As commented in \cite{Girelli:2006fw}, Finsler geometry could be used to describe both breakdown and deformation of spacetime symmetries. However we have showed here that this symmetry deformation implies the necessity to characterize velocities as living in a tangent space with non-trivial translation laws. Those laws moreover result to be compatible with the phase-space deformation of the theory (\ref{FlatHamilt}).

\section{Conclusions}
The question of what may be the nature of the spacetime emerging from a semi-classical limit of a quantum description of the gravitational degrees of freedom is still open. Possibly, we may have a dimensional reduction coming from CDT, Ho\v{r}ava-Lifshitz gravity and others (see \cite{Amelino-Camelia:2013tla} and references therein), non-local metrics \cite{Padmanabhan:2015vma}, Rainbow-like metrics coming from a quantization of geometrical and matter-like degrees of freedom \cite{Assaniousssi:2014ota} and, as we have shown, a possible Finsler nature coming from a MDR. In all these cases there exists a common point:  the non-Riemannian nature of spacetime. Which is not a complete surprise, after all there is no {\it a priori} request for nature of spacetime to be limited to Riemannian geometry at all scales, despite the abundance of well-defined geometrical structures (not to mention those that are still to be discovered).
\par
In this paper, we analyzed the possibility of having a deformed dispersion relation within a curved background, investigating what kind of manifold could be the most natural to ``host'' them. We generalized the results presented in \cite{Girelli:2006fw,Amelino-Camelia:2014rga} and we proposed a solution for the massless limit problem, {\it id est} we found a candidate partially-homogeneous metric using, which are possible to formalize, the most important properties that a proper geometrical model for the description of elementary particles should provide, such as the geodesic equation and the dispersion relation. The presence of such unified Finsler structure that an observer would assign to spacetime, when performing measurements observing high-energetic particles, allowed us to formalize the usual phenomenology of time delays of photons from GRBs due to the energy-dependent speed of light presented in \cite{Jacob:2008bw}, plus a correction of the order of the Planck length originated from the non-Riemannian structure of spacetime, with these both contributions having the same geometrical origin. One may also wonder whether the possibility of expressing some Planck-scale-deformed Hamiltonian formalism as a Finsler geometry is an artifact of the linearization with respect to $\ell$. We know for sure that at first order in the deformation parameter this formalization is exact, however we have no elements to state that this should also be the case at all orders. Then in order to investigate on such an exact theory, one may need to further generalize the Finsler picture. However since our investigation relies on the Legendre transform to switch from the Hamiltonian formalism to the Lagrangian one, an all-orders exploration is now not possible in general and we should leave this concern to further investigations.
\par
We estimated the correction that such a model could bring to the cosmological redshift effect. In this context, it would be extremely interesting to test such effect using cosmological data, however unfortunately the best observations on some kind of energy dependence for the cosmological redshift yet published \cite{Ferreras:2016xsq} do not have the sensitivity to discriminate the kind of effect we found here. The authors found no energy dependence with a precision of $\Delta z\sim 10^{-6}$, whereas in the visible range of electromagnetic radiation ($\sim 5000 \text{\AA}$) the effect we take into account would be of the order of $\Delta z\sim 10^{-28}$ on the redshift magnitude. In any case, we need phenomenological analysis with more energetic particles.
\par
The last open issue we explored is the deformation of velocities composition law in the Finsler tangent space that symmetrical $\ell$-deformed theories require (as already pointed out in \cite{Amelino-Camelia:2014rga}), in order to allow to different observers to agree on the occurrence (or not) of some interaction event. We found that introducing a slightly more complicated formalism inspired by Hopf algebras literature it is still possible to obtain a full relativistic description for particles interactions.

\subsection*{Acknowledgements}

NL acknowledges support by the European Union Seventh Framework Programme (FP7 2007-2013) under grant agreement 291823 Marie Curie FP7-PEOPLE-2011-COFUND (The new International Fellowship Mobility Programme for Experienced Researchers in Croatia - NEWFELPRO), and also partial support by the H2020 Twinning project n$^\text{o}$ 692194, RBI-TWINNING. 
\par
IPL is supported by the International Cooperation Program CAPES-ICRANet financed by CAPES - Brazilian Federal Agency for Support and Evaluation of Graduate Education within the Ministry of Education of Brazil grant BEX 14632/13-6.
\par
FN acknowledges support from CONACYT grant No. 250298. 

%%%%%%%%%%%%%%%%%%%%%%%%%%%%%%%%%%%%%%%%%%%%%%%%%%%%%%%%%%%%%%%%%%%%%%%%%%%%%%%%%%%%%%%%%

\appendix

\section{Connection with Relative Locality}

First of all let us define a generic\footnote{This one of course is not the most generic momentum-space metric one can possibly define, however for the purposes of this article we will not need any further parametrization.} momentum-space metric depending on our two parameters $\beta,\gamma$, at first order in the deformation parameter $\ell$:
\begin{equation}
\zeta_\ell^{\mu\nu}(P)=\left(\begin{array}{cc}
1+2 \gamma\ell P_0  & 0\\
0 & -e^{-2 H x^0}(1-2\beta\ell P_0)
\end{array}\right)\,.
\end{equation}
As well established in literature \cite{AmelinoCamelia:2011nt,Loret:2014uia,Amelino-Camelia:2013uya} we can find the Hamiltonian by integrating the momentum-space invariant line-element
\begin{equation}
{\cal D}p=\int_0^1 ds\sqrt{\zeta^{\alpha\beta}(P)\dot{P}_\alpha\dot{P}_\beta}\,,\label{intCasimir}
\end{equation}
where the momentum-space geodesics $P(s)$ are defined by the geodesic equation: 
\begin{equation}
\ddot{P}_\alpha + \Gamma_\alpha^{\beta\gamma}\dot{P}_\beta\dot{P}_\gamma= 0\,,
\end{equation}
in which the connections are our usual momentum-space Christoffel symbols: 
\begin{equation}
\Gamma_\lambda^{\mu\nu}=\frac{1}{2}\left(\frac{\partial}{\partial p_\mu}\zeta^{\sigma\nu}+\frac{\partial}{\partial p_\nu}\zeta^{\sigma\mu}-\frac{\partial}{\partial p_\sigma}\zeta^{\mu\nu}\right)\,.
\end{equation}
Given our generic metric, then, we just have to solve the following equations:
\begin{eqnarray}
&&\ddot{P}_0 + \ell\left(\gamma\dot{P}_0^2-\beta e^{-2 H x^0} \dot{P}_1^2)\right)= 0\,,\nonumber\\
&&\ddot{P}_1 - 2\ell\beta\dot{P}_0\dot{P}_1= 0\,.
\end{eqnarray}
Since $\dddot{P}\sim O(\ell^2)$, we observe that in our case the geodesic can be expressed as
\begin{equation}
P_\alpha(s)\simeq \bar{p}_\alpha + s \dot{P}_\alpha |_{s=0}+\frac{1}{2}s^2 \ddot{P}_\alpha |_{s=0}\,.
\end{equation}
Let us now impose the initial values for the geodesic as $P_\alpha(0)=0$ and $P_\alpha(1)=p_\alpha$. It is easy to notice that they imply $\bar{p}_\alpha=0$ and $p_\alpha=\dot{P}_\alpha |_{s=0} +\frac{1}{2} \ddot{P}_\alpha |_{s=0}$.
We have now all we need in order to solve our equations, {\it id est}:
\begin{eqnarray}
\dot{P}_0 |_{s=0}&=& p_0 +\frac{\ell}{2}\left(\gamma p_0^2  -\beta e^{-2 H x^0} p_1^2\right)\,,\nonumber\\
\dot{P}_1 |_{s=0}&=& p_1 - \ell \beta p_0 p_1\,,
\end{eqnarray}
and also
\begin{eqnarray}
\ddot{P}_0 |_{s=0}&=& -\ell\left(\gamma p_0^2 -\beta e^{-2 H x^0} p_1^2\right)\,,\nonumber\\
\ddot{P}_1 |_{s=0}&=& + 2\ell \beta p_0 p_1\,.
\end{eqnarray}
Therefore we are now able to solve integral (\ref{intCasimir}), and find the generic Hamiltonian expression:
\begin{equation}
{\cal H}= p_0^2-p_1^2 e^{-2 H x^0}+\ell\left(\gamma p_0^3+\beta p_0 p_1^2 e^{-2 H x^0}\right)\,.\label{CasimirParametric}
\end{equation}

%%%%%%%%%%%%%%%%%%%%%%%%%%%%%%%%%%%%%%%%%%%%%%%%%%%%%%%%%%%%%%%%%%%%%%%%%%%%%%%%%%%%%%%

\section{Solving the Killing equations in the flat-spacetime limit}

In the introductory section we have shown that the Lagrangian-Hamiltonian Legendre transformation defines the relation
\begin{equation}\label{b1}
\zeta_{\mu\nu}\dot{x}^\mu\dot{x}^\nu=2 p_\alpha\dot{x}^\alpha-\tilde{g}_{\rho\sigma}\dot{x}^\rho\dot{x}^\sigma\,.
\end{equation}
Since eq. (\ref{Cxpunto}) basically states that the first term of \ref{b1} must be invariant under boost, then we have that
\begin{equation}
\delta\left(2 p_\alpha\dot{x}^\alpha-\tilde{g}_{\rho\sigma}\dot{x}^\rho\dot{x}^\sigma\right)=\{{\cal N},{\cal H}(\dot{x})\}=0\,.
\end{equation} 
Therefore
\begin{equation}
\delta(\tilde{g}_{\mu\nu}\dot{x}^\mu\dot{x}^\nu)=2 p_\alpha \delta(\dot{x}^\alpha)\,.
\end{equation}
However as already noticed in \cite{Amelino-Camelia:2014rga}, being spacetime flat we have $\dot{p}_\mu=0$, and then the second term of last equation is just a total derivative:
\begin{equation}
p_\mu\delta\dot{x}^\mu=\frac{d}{d\tau}(p_\mu\delta x^\mu)=\frac{d}{d\tau}\left(\frac{\partial {\cal L}}{\partial\dot{x}^\mu}\delta x^\mu\right)\,,
\end{equation}
which can be eliminated during the integration procedure. The Killing equation of our theory is then equivalent to (\ref{KillingEq}), of course in the flat spacetime limit, being
\begin{equation}
\delta(\tilde{g}_{\mu\nu} dx^\mu dx^\nu)=0\,,
\end{equation}
excepted total derivative terms.\\
Given this, from Killing equation (\ref{KillingEq}) we find the system of differential equations we need to solve in order to find our symmetry generators:
\begin{equation}
\left\{\begin{array}{l}
\tilde{g}_{0\mu}\partial_0\xi^\mu+
\frac{1}{2}\frac{\partial \tilde{g}_{00}}{\partial\dot{x}^0}\partial_\alpha\xi^0\dot{x}^\alpha =0\\
\\
\tilde{g}_{1\mu}\partial_1\xi^\mu+
\frac{1}{2}\frac{\partial \tilde{g}_{11}}{\partial\dot{x}^0}\partial_\alpha\xi^0\dot{x}^\alpha =0\\
\\
\tilde{g}_{\mu 1}\partial_0\xi^\mu+\tilde{g}_{0\nu}\partial_1\xi^\nu+\frac{\partial \tilde{g}_{01}}{\partial\dot{x}^1}\partial_\alpha\xi^1\dot{x}^\alpha=0\\
\end{array}\right.\,.\label{eq:killsysgenerico}
\end{equation}
We look for perturbative solutions at first order in $\ell$ such as
\begin{equation}
\xi^0=\xi^0_{(0)}+\ell\xi^0_{(\ell)}\;,\;\;\;\xi^1=\xi^1_{(0)}+\ell\xi^1_{(\ell)}\,,
\end{equation}
in which the 0th order solutions are the Minkowskian ones:
\begin{equation}
\xi^0_{(0)}=c^0+k x^1\;,\;\;\;\xi^1_{(0)}=c^1 + k x^0\,.
\end{equation}
 This leads to
\begin{equation}
\left\{\begin{array}{l}
\partial_0\xi^0_{\ell}=\left(\frac{\gamma}{2}+\beta_2\right) k \dot{x}^1\\
\\
\partial_1\xi^1_{(\ell)}=-\left(\beta_2+\frac{\beta_1}{2}\right) k \dot{x}^1\\
\\
\partial_0\xi^1_{(\ell)}-\partial_1\xi^0_{(\ell)}=-(\beta_1+\beta_2+\gamma) k \dot{x}^0\\
\end{array}\right.\,.\label{eq:killsysparam}
\end{equation}

From system (\ref{eq:killsysparam}) first and second equation we just obtain that the generalized Killing vector components should have the form:
\begin{equation}
\xi^0_{(\ell)}=\kappa^0(x^1,\dot{x})+\left(\frac{\gamma}{2}+\beta_2\right)\dot{x}^1 x^0\;,\;\;\;\xi^1_{(\ell)}=\kappa^1(x^0,\dot{x})-\left(\beta_2+\frac{\beta_1}{2}\right)\dot{x}^1 x^1.\label{Kill1}
\end{equation}
Now using the results obtained (\ref{Kill1}) with the third equation, re-deriving this one we get
\begin{eqnarray}
\kappa^0(x^1,\dot{x})&=&\mathbb{C}^0(\dot{x})+\kappa^0(\dot{x})x^1\,,\\
\kappa^1(x^0,\dot{x})&=&\mathbb{C}^1(\dot{x})+\kappa^1(\dot{x})x^0\,.\label{xi1imperfetta}
\end{eqnarray}
Then, using (\ref{xi1imperfetta}) again with the third equation of system (\ref{eq:killsysparam}) we obtain the relation between $\kappa^0$ and $\kappa^1$ which is:
\begin{equation}
\kappa^0(\dot{x})=\kappa^1(\dot{x})+(\beta_1+\beta_2+\gamma) k\dot{x}^0\,.
\end{equation}
 We can now express the complete solutions for the $\xi^\alpha_{(\ell)}$ as
\begin{eqnarray}
\xi_{(\ell)}^0&=&\mathbb{C}^0(\dot{x})+\kappa^1(\dot{x})x^1+k\left((\beta_1+\beta_2+\gamma)\dot{x}^0 x^1+\left(\frac{\gamma}{2}+\beta_2\right)\dot{x}^1 x^0\right)\,,\\
\xi_{(\ell)}^1&=&\mathbb{C}^1(\dot{x})+\kappa^1(\dot{x})x^0-k\left(\beta_2+\frac{\beta_1}{2}\right)\dot{x}^1 x^1\,.
\end{eqnarray}
The natural choice $\mathbb{C}^0(\dot{x})=\mathbb{C}^1(\dot{x})=\kappa^1(\dot{x})=0$ gives us the total charge, in fact it is easy to verify that in this case:
\begin{eqnarray}
Q_\ell&=&\xi^\mu p_\mu= c^0 p_0+c^1 p_1 + 
k\left(x^1 p_0+(\beta_1+\beta_2+\gamma)\ell x^1 \dot{x}^0 p_0-\frac{\ell}{2}(\beta_1+2\beta_2)x^1 \dot{x} p_1 + x^0 p_1+\ell\left(\frac{\gamma}{2}+\beta_2\right) x^0 \dot{x}^1 p_0\right)\simeq\nonumber\\
&\simeq& c^0 p_0+c^1 p_1 +k\left( x^0 p_1\left(1-\ell\left(\beta_2+\frac{\gamma}{2}\right) p_0\right) + x^1\left(p_0+(\beta_1+\beta_2+\gamma)\ell p_0^2+\frac{\ell}{2}(\beta_1+2\beta_2)p_1^2\right)\right).
\end{eqnarray}
The choice $\kappa^1(\dot{x})=0$ works fine in the diagonal case metric. However Finsler formalism relies on homogeneous metrics, which in general have non zero off-diagonal terms. The total charge general expression imposing no particular condition to our parameters at first order in $\ell$ has the form
\begin{equation}
Q_\ell=\xi^\mu p_\mu \simeq (c^0+\ell{\mathbb C}^0(\dot{x}))p_0 + (c^1+\ell\mathbb{C}^1(\dot{x}))p_1 + (k + \ell\kappa^1(\dot{x})){\cal N}\,,
\end{equation}
in which $p_0$, $p_1$ and $\mathcal{N}$ are the deformed-Poincar\'{e} algebra generators in 1+1D. In order to obtain a coherent physical framework ({\it id est} invariant Casimir and particles worldlines) also in the Finsler case, we need then to impose $\kappa^1(\dot{x})=\left(\beta_2-\frac{\gamma}{2}\right)\dot{x}^0$. This choice eliminates any direct sign of the metrics off-diagonal terms in the boost representation (\ref{BoostRepresentation}).

\end{document}